\documentclass[12pt]{article}
\usepackage[margin=1 in]{geometry}
\usepackage{amsmath}
\usepackage{enumerate}
\usepackage{natbib}
\usepackage{caption}
\usepackage{subcaption}
\usepackage{xcolor}
\usepackage{graphicx,psfrag,epsf}
\usepackage{setspace}
\usepackage{url} 

\newcommand{\blind}{0}

\def\bSig\mathbf{\Sigma}

\begin{document}

\def\spacingset#1{\renewcommand{\baselinestretch}%
{#1}\small\normalsize} \spacingset{1}
\doublespacing


\if0\blind
{
  \title{\bf Desirability of outcome ranking (DOOR) analysis for multivariate survival outcomes with application to ACTT-1 trial}
  \author{Shiyu Shu, Guoqing Diao\thanks{
    Contact gdiao@email.gwu.edu
    }\hspace{.2cm}, Toshimitsu Hamasaki, and Scott Evans\\
    Department of Biostatistics and Bioinformatics\\
    The George Washington University}
    \date{}
  \maketitle
} \fi

\if1\blind
{
  \bigskip
  \bigskip
  \bigskip
  \begin{center}
    {\LARGE\bf Connecting longitudinal desirability of outcome ranking (DOOR) analysis with multivariate survival outcomes}
\end{center}
  \medskip
} \fi

\bigskip
\begin{abstract}
Desirability Of Outcome Ranking (DOOR) methodology accounts for problems that conventional benefit:risk analyses in clinical trials ignore, such as competing risks and the trade-off relationship between efficacy and toxicity. DOOR levels can be considered as a multi-state process in nature, as event-free survival, and survival with side effects are not equivalent and the overall patient trajectory requires recognition. In monotone settings where patients’ conditions can only decline, we can record event times for each transition from one level of the DOOR to another, and construct Kaplan-Meier curves displaying transition times. While traditional survival analysis methods such as the Cox model require assumptions like proportional hazards and suffer from the  challenge of interpreting a hazard ratio, Restricted Mean Survival Time (RMST) offers an alternative with greater intuitiveness. Therefore, we propose a combination of the two domains to develop estimation and inferential procedures that could benefit from the advantages of both DOOR and RMST. Particularly, the area under each survival curve restricted to a time point, or the RMST, has clear clinical meanings, from expected event-free survival time, expected survival time with at most one of the events, to expected lifetime before death. We show that the nonparametric estimator of the RMSTs asymptotically follows a multivariate Gaussian process through the martingale theory and functional delta method. There are alternative approaches to hypothesis testing that recognize when patients transition into worse states. We evaluate our proposed method with data simulated under a multistate model. We consider various scenarios,  including when the null hypothesis is true, when the treatment difference exists only in certain DOOR levels, and small-sample studies. We also present a real-world example with ACTT-1.
\end{abstract}

\noindent%
{\it Keywords:}  clinical trial, competing risks data, influence function, martingale theory, restricted mean survival time
\vfill

\newpage
\spacingset{1.45} 
\section{Introduction}
\label{sec:intro}

Competing risks in clinical trials could often blur our inference about the treatment effect, and a traditional Cox proportional hazard model will lead to biased results under such scenario \citep{b1}. In the last 15 years, a family of nonparametric methods that utilize Wilcoxon-Mann-Whitney(WMW) U statistics is gaining popularity \citep{b2, b3}. Namely, \cite{b4} proposed Generalized Pairwise Comparison to analyze Proportion in Favor of Treatment/Net Treatment Benefit/Somers' D, \cite{b5} proposed sequential comparison algorithms to estimate Win Ratio(WR) and \cite{b8} proposed Desirability of Outcome Ranking(DOOR) to estimate DOOR probability, all recognizing the unequal importance between multiple outcomes and evaluating the treatment effect through pairwise comparisons at patient level \citep{b7}. Suppose in a two-arm clinical trial, we have $n_1$ participants from one group and $n_2$ participants from the other. We can form $n_1n_2$ many patient-pair comparisons in total, and set up rules to determine the win, loss or tie of each pair at the individual level, and make inferences about whether a randomly selected patient in one group could have a more desirable outcome than a randomly selected patient in the other group on the population level. In such a fashion, we utilize outcomes to analyze patients rather than patients to analyze outcomes \citep{b6}.\\

For example, the DOOR methodology ranks patients' outcomes based on clinical importance and desirability, and the construction of DOOR outcomes requires tailoring to the specific disease area \citep{b6, b8}. The methodology has received much attention in the medical world, where \cite{b9} defines an 8-level DOOR for an obstetrical trial, and \cite{b10} applies a 5-level DOOR to Complicated Urinary Tract Infection. ACTT-1 \citep{b14} is a clinical trial that compares the treatment effect of Remdesivir and placebo on COVID symptoms, where patient conditions over 8 visits in a month were recorded. While analyzing any single endpoint at Day 29 marginally would fail to reject the null, the DOOR probability indicates that the treatment effect from Remdesivir is statistically significantly better than placebo. In an ad-hoc analysis, we define a four-level DOOR as below (from most desirable to least desirable):\\

\begin{center}
1. Alive without hospitalization or serious adverse events(SAE) \\
2. Alive with either hospitalization or SAE \\
3. Alive with both hospitalization and SAE \\
4. Death \\
\end{center}

However, these methods have two limitations: First, if the endpoints are indeed time-to-event, and we observe ties in ranks between a pair, we often use time-to-event as tie-breakers \citep{b4,b5,b8}. It is important to note that relative time, i.e., the order of occurrence of events, is used as a tie-breaker when the ranks are tied between a patient-pair comparison, but not the exact time, and we could lose some information. Suppose patient A and patient B receive placebo and treatment, respectively, and both experience event E. Regardless of whether the days to event E are 1 vs. 2, or 1 vs. 100, treatment will be considered as a win over placebo, but the two scenarios will have totally different clinical meaning. Second, the summary measure of these methods, e.g., proportion in favor of treatment, win ratio or DOOR probability, explains the treatment effect on a probability/ratio scale, and is thus hard to interpret for clinicians. As \cite{b5} mentioned in the discussion, the magnitude of the win remains a vague idea.\\

On the other hand, Cox models \citep{b11} in survival analysis also suffer from the lack of explainability of hazard ratios, and restricted mean survival time(RMST) translates the survival problem to the actual time scale \citep{b12}. RMST is the expected survival time up until a restricted time threshold $\tau$, and can be estimated with both parametric and nonparametric approaches. In particular, RMST can be estimated nonparametrically as the area under  a Kaplan-Meier curve \citep{b13}, and its associated variance can be estimated in various forms \citep{b15, b16, b17}.\\

Researchers have applied the idea of RMST in many directions: \cite{b19} extends the setting to observational studies; \cite{b20} describes a meta-analysis procedure to estimate the standard deviation of RMST; \cite{b21} addresses asymptotic properties on the choices of time; and \cite{b22} includes covariates to predict RMST. This concept is also widely embraced by investigators and has been applied in oncology studies \citep{b23, b24}.\\

In this paper, we would like to combine the merits of DOOR and RMST, where we generate tiered Kaplan-Meier curves based on the levels of DOOR, estimate the covariance structure of multiple RMSTs, and make inferences within and across treatment arms. As competing risk is a common challenge in survival analysis, previous works in the RMST literature \citep{b25, b26, b27} proposed potential corrections through inverse probability weighting (IPW) or cumulative incidence function (CIF). Our hypothesis testing procedure shares similarities to procedures described in \cite{b26}, but benefits from the construction of DOOR, and might provide further clinical insights.\\

In Section 2, we describe the methods; Section 3 provides simulation results based on a hypothetical scenario that mimics real-world settings, and an application to the motivating example ACTT-1 is presented in Section 4. We conclude the paper with discussions in Section 5. \\ 

\section{Methods}
\label{sec:meth}


The methodology originates from the definition of DOOR levels, and without loss of generality, we assume the following 5-level scenario:
\begin{center}
1. Survived with 0 events\\
2. Survived with 1 event w/o disability\\
3. Survived with $>$1 event w/o disability\\
4. Survived with overall disability\\
5. Death\\
\end{center}

The qualifying events include:
\begin{center}
1. Stroke\\
2. Severe bleeding\\
3. Disability (implies category 4)\\
4. Death (implies category 5)\\
\end{center}

With five levels of DOOR, we can define four gaps as the time to the next/worse level. Formally, we define the following event times:
\begin{center}
1. $T_1$ =  event-free survival time, or time to the first occurrence of any qualifying events\\
2. $T_2$ = time to both stroke and severe bleeding, or disability, or death\\
3. $T_3$= time to disability\\
4. $T_4$ = time to death\\
\end{center}

The event times will also have corresponding binary censoring indicators $D_1, D_2, D_3, D_4$. Since the construction of DOOR is ordinal, it is natural that $T_1\leq T_2 \leq T_3 \leq T_4$. Figure 1 gives a graphical representation of potential scenarios of patient disease progression.\\

We denote the restricted time as $\tau$ and censoring time $C\leq \tau$. For each type of event time $T_j ~~ (j=1,...,4)$ defined above, we can estimate the survival function $S_j(t)$ using the Kaplan-Meier estimator $\widehat{S}_j(t)$.  Define the restricted mean survival time up to time $\tau$ as $\phi(S_j(\cdot); \tau) = \int_0^\tau S_j(s) ds=E[min(T_j, \tau)]$, which is the area under the survival curve $S_j(\cdot)$ up to time $\tau$.  We further define: (1).$Y_j(s)=\sum_{i=1}^nY_{ij}(s)$, $Y_{ij}(s)$ indicates the survival status for event $j$ of patient $i$ at time $s$, and $Y_j(s)$ is the number of patients at risk; (2). proportion of patient at risk, $\pi_j(t) = P(\min\{T_j, C\} \geq t)$, which can be estimated by $Y_j(t)/n$, where n is the total sample size; (3). $A_j(t)$, the cumulative hazard function for the $j$th type event, and $\alpha_j(t)$ the hazard function; (4).$N_j(t)=\sum_{i=1}^nI(\min\{T_{ij}, C\} \leq t, D_{ij}=1)$, the counting process for the $j$th type event; (5). $M_j(t)=N_j(t)-\int^t_0Y_j(s)dA_j(s)$, the zero-mean martingale typically used in survival analysis for the counting process $N_j(t)$, where the second term $\Lambda_j(t)=\int_0^tY_j(s)\alpha_j(s)ds$ represents the cumulative intensity process, or compensator process \citep{b28}.\\

There is a vast literature in survival analysis on restricted mean survival time, martingale theory, and Nelson-Aalen estimator, so we will not go into the mathematical details of the proofs, and will instead use existing theorems. We establish the following asymptotic properties for the estimators of multiple RMSTs:\\

{\bf Theorem} For any $\tau \in \{s: \pi_j(s) > 0\}$, $\phi(\widehat{S}_j(\cdot); \tau)=\int_0^\tau \widehat{S}_j(t) dt$ is the estimator of $\phi(S_j(\cdot); \tau)$. Asymptotically, \[
n^{1/2}\{\phi(\widehat{S}_j(\cdot); \tau) - \phi({S}_j(\cdot); \tau)\} \rightarrow 
W\{\zeta_j(\tau)\},
\] in distribution on $[0,\tau]$ almost surely, where $W$ is Brownian motion and $\zeta_j$ will be defined later in Appendix.\[
\begin{split}
& \text{Cov}[n^{1/2}\{\phi(\widehat{S}_j(\cdot); \tau) - \phi({S}_j(\cdot); \tau)\},
n^{1/2}\{\phi(\widehat{S}_k(\cdot); \tau) - \phi({S}_k(\cdot); \tau)\} ]  \\
&\rightarrow_P \text{Cov}\{ \Psi_{ij}(\tau), \Psi_{ik}(\tau)\},\\
\end{split}
\] where $\Psi_{ij}(\tau)$ is the individual influence function/canonical gradient \citep{b29} that will be defined later in the Appendix, along with proofs of the theorem.\\

Once we obtain the estimators of the RMSTs $\hat R_T, \hat R_C$ respectively for the treatment arm and the control arm, and the covariance matrix estimators $\hat\Sigma_T, \hat\Sigma_C$. We can propose four different hypothesis testing:
\begin{center}
1. Inference about a single RMST, $\widehat{var}(\hat R_{Tj})=\hat\Sigma_{Tjj}$\\
2. Inference about the difference of two RMSTs between two arms, $\widehat{var}(\hat R_{Tj}-\hat R_{Cj})=\hat\Sigma_{Tjj}+\hat\Sigma_{Cjj}$\\
3. Inference about the difference of two RMSTs within the same arm, $\widehat{var}(\hat R_{Tj}-\hat R_{Tk})=\hat\Sigma_{Tjj}+\hat\Sigma_{Tkk}-2\hat\Sigma_{Tkj}$\\
4. Inference about the overall equivalence of the two sets of trajectories, which leads to a wald test with Chi-sq statistics $(\hat R_{T}-\hat R_{C})^T(\hat\Sigma_{T}+\hat\Sigma_{C})^{-1}(\hat R_{T}-\hat R_{C})$, with df equal to the types of RMST\\
\end{center}

\section{Results}
\label{sec:resu}
\subsection{Simulations}

We conduct simulation studies to examine the finite-sample performance of the proposed methodology. Without loss of generality, we follow the setup described in Figure 1, and generate event times $T_1, T_2, T_3, T_4$ under a multi-state model. Among five possible states(I, initial state; SB1, first stroke or bleeding; SB2, second stroke or bleeding; DI, disability; DE, death), we have nine possible transitions: $I \to \{SB1, DI, DE\}, SB1 \to \{SB2, DI, DE\}, SB2 \to \{DI, DE\}, DI \to DE$.\\

We code the above five states from I to DE with numbers 1 to 5, respectively. Let $\lambda_{jk}(t)$ be the transition hazard from state $j$ to $k\in A(j)$, where $A(j)$ contains all possible states that state $j$ can transit to, and $\lambda_{j}(t)=\sum_{k\in A(j)}\lambda_{jk}(t)$. The data generation is as follows:\\

1. Generate $T_1$ from a distribution with hazard $\lambda_1(t)$, where $A(1)=\{2,4,5\}$. The event type $D_1$ is generated from $P(D_1=j)=\lambda_{1j}(T_1)/\lambda_1(T_1), j=2,4,5$. If $D_1=5$(death), then $T_j=T_1, j=2,3,4$. If $D_1=4$(disability), then $T_j=T_1, j=2,3$.\\

2. If $D_1=2$(first stroke or bleeding), we generate gap time between $T_1$ and $T_2$, denoted by $G_2$, from a distribution with hazard $\lambda_2(t)$, where $A(2)=\{3,4,5\}$. For the second event $D_2$, we can analogously generate $P(D_2=j), j=3,4,5$. If $D_2=5$, $T_4=T_3=T_2+T_1+G_2$. If $D_2=4$, $T_3=T_2=T_1+G_2$.\\

3. If $D_2=3$(second stroke or bleeding), we generate gap time $G_3$ from $\lambda_3(t), A(3)=\{4,5\}$. For the third event $D_3$ generated from $P(D_3=j)$, $T_4=T_3=T_2+G_3$ if $D_3=5$; $T_3=T_2+G_3$ if $D_3=4$.\\

4. If $D_1=4$, or $D_1=2 \& D_2=4$, or $D_1=2 \& D_2=3 \& D_3=4$, we generate gap time between $T_3$ and $T_4$.\\

5. We generate censoring time $C$ from a uniform distribution. The indicator for event $j$ will be $0$ if $C<T_j$, and $1$ if $C\geq T_j$. The time-to-event j is the minimum of $T_j$ and $C$.\\

Following this sequential algorithm, we restrict the occurrence of a more severe event from before the occurrence of a less severe event, reflecting the ordinal nature of DOOR. Note that the transition rates and distribution of true event times are not of primary interest here, as long as the simulated data has the ordinal structure we desire, so we only consider one combination of transition rates, and generate true event times from an exponential distribution. The true RMST values may not have a closed-form solution due to the introduction of censoring, so we approximate the true values through a million Monte Carlo simulation.\\

Based on our choice of censoring distribution $Unif(0,4)$, we consider three different choices(1, 1.5, or 2) for $\tau$, the time window, and conduct simulation under 1000 replicates under two sample sizes(100 or 400). We report bias, standard deviation of the estimate(SE), standard error estimate(SEE), coverage rates of 95\% confidence intervals(CP), and the average number of corresponding events in Table \ref{t1}.\\

From Table \ref{t1}, we obtain results as expected. Across all scenarios, the bias between the estimate of true RMST values is relatively small, and SE is approximately the same as SEE. We observe CP below the nominal coverage of 95\% in a few cases, which is due to the low number of events when the sample size is small or the time window is early. With 400 observations per replicate, we could achieve asymptotic results.\\

Additionally, we evaluate the power of our method by making type II and type IV inferences mentioned in the Methods section, as hypothesis testing for the other two types is of less interest. We consider the transition rates to be $(0.3, 0.15, 0.06, 0.6, 0.3, 0.12, 0.36, 0.24, 0.24)$ under the null, and $(0.5, 0.2, 0.1, 1, 0.4, 0.2, 0.6, 0.3, 0.3)$ under the alternative. Under the alternative at $\tau=2$, for example, the true RMSTs for the placebo arm will be $0.998, 1.247, 1.392, \\ 1.726$ respectively, and $1.25, 1.49, 1.58, 1.83$ for the treatment arm.\\

From Figure \ref{fig2}, we can see that if we want to test the equivalence of individual-level RMST between arms, the type I error is controlled at the nominal level at 0.05, while the power increases as the sample size and/or time increases. Due to the tiered nature of our RMSTs, the event type corresponding to $RMST_1$ will always have more events than later event types, so we will have a higher power of detecting the difference in $RMST_1$, analogous to the trend we observe in the coverage probability.\\

From Figure \ref{fig3}, the power to reject the null is higher under the alternative because we pool all events in each arm together and test whether the entire trajectory patterns are the same between arms.\\

\subsection{Real World Example - ACTT-1 Trial}

Adaptive COVID-19 Treatment Trial (ACTT-1) is a multicenter, adaptive, randomized blinded controlled trial of the safety and efficacy of investigational therapeutics for the treatment of COVID-19 in hospitalized adults. There was no known efficacious treatment for COVID-19 at the time. A total of 1,062 Patients were randomly assigned to receive either Remdesivir (541 patients, 200 mg loading dose on day 1, followed by 100 mg daily for up to 9 additional days) or placebo (521 patients) for up to 10 days. All patients were followed for up to 8 times in a month.\\

Three endpoints were considered in the study: 1. Hospitalized with invasive mechanical ventilation/ECMO 2. Serious Adverse Events (SAE) 3. Death. We thus define three time points and their corresponding RMSTs: time to occurrence of first qualifying event, $RMST_1$=Event-free survival; time to both SAE and Hospitalization, or death, $RMST_2$=survival with either event; time to death, $RMST_3$=survival with both events. To ensure our data is suitable for survival analysis, we restrict the patient outcome to be the least desirable condition observed since randomization for monotonicity. Figure \ref{fig4} visually presents the ordinal Kaplan-Meier curves, and we could see some differences in the patterns.\\

In order to quantify the difference we see in the visualization, we set $\tau=20$ and summarize the estimated RMSTs in Table \ref{t2}. If we combine the type II inferences summarized in Table \ref{t2} and Figure \ref{fig5}, we observe that patients receiving Remdesivir, on average, have approximately two more event-free days than patients receiving placebo, which could be clinically important. At 20 days after randomization, Remdesivir performs better than Placebo in preventing patients from experiencing any events and preventing future events from occurring when patients already experience any events. The efficacy of Remdesivir reflected in RMSTs is consistent with clinical findings in \cite{b30} and findings in \cite{shu2024longitudinal}, where a longitudinal DOOR endpoint was considered. The light green area in Figure \ref{fig4} represents the difference between $RMST_1$ and $RMST_2$, and has an intuitive clinical meaning: the expected time that a patient stays in the state of having one event. (Type III inference) In the treatment arm, the difference is 5.17 days, with a 95\% confidence interval (4.54, 5.80).
For the general Wald test, we have test statistic $18.05>7.81$, the critical value for the Chi-sq distribution with 3 degrees of freedom, so we reject the null hypothesis that two RMST sets are equivalent. (Type IV inference)\\

\section{Discussion}
\label{sec:disc}

In this paper, we introduce the RMST DOOR methodology, provide asymptotic properties of multi-level RMST estimators, present different ways to make inference that addresses different clinical questions, demonstrate the effectiveness of our method through simulation, and give an example of analyzing clinical trial data. Most clinical trials collect information on multiple outcomes that could be competing risks, and our extension originating from the ordinal nature of DOOR construction allows for a comprehensive benefit:risk analysis, as well as an intuitive measure for clinicians to interpret.\\

Different groups have addressed the analysis of multiple time-to-event endpoints from different angles. \cite{b27} analyzes the same ACTT-1 data by jointly making inference about time gained due to hospital discharge and time lost due to death, and \cite{b31} proposes a trial design with win ratio ideas. Our proposed method differs in first having the construction of an ordinal DOOR outcome in mind, where various endpoints such as hospitalization, SAE, and death can be combined in a meaningful way, and analyzing patient outcomes in a cumulative nature, rather than comparing outcomes in a hierarchical algorithm as in win ratio. Unlike traditional RMST analysis with a single outcome, which is typically time to death, we consider a wider range of outcomes and the analysis is rather RMET (Restricted Mean Event Time), and may increase statistical power by incorporating the totality of the evidence.\\

We hope that this method could expand current research in survival analysis and benefit:risk analysis by providing a tool to utilize the totality of evidence. Possible future directions include applications to other disease areas such as cardiovascular complications in diabetes or cancer, simultaneous confidence bands of the RMSTS at different $\tau$, incorporation of survival probability estimators other than KM (e.g., Nelson-Aalen or weighted KM), estimation of RMSTs with covariate adjustments, and development of R packages and shiny app.\\

\section{Conclusion}

Restricted mean survival time is a useful tool to analyze treatment effect nonparametrically. By combining the merits of RMST and DOOR, we offer a framework that could make inference on multiple RMSTs defined in an ordinal structure, which enables researchers to address various clinical questions.\\

\section*{Acknowledgment}
\label{sec:ackn}

The authors report there are no competing interests to declare.

\section*{Appendix}

{\bf Proof}: Following an application of Duhamel's equation in survival analysis(equation 9 in \cite{b32}, equation 66 in     \cite{b33}), we have
\[
\begin{split} 
\widehat{S}_j(\tau)
&=S_j(\tau)-S_j(\tau)\int^\tau_0\widehat{S}_j(s-)/S(s)*dM_j(s)/Y_j(s)\\
&\approx S_j(\tau)-S_j(\tau)\int^\tau_0Y^{-1}_j(s)dM_j(s).\\
\end{split}
\]

Then we can integrate the equation on both sides, and get 
\[
\begin{split}
\phi(\widehat{S}_j(\cdot); \tau)
&=\int^\tau_0[S_j(t)-S_j(t)\int^t_0I(s<t)Y^{-1}_j(s)dM_j(s)]dt\\
&=\phi({S}_j(\cdot); \tau)-\int^\tau_0[\int^\tau_0I(s<t)S_j(t)dt]Y^{-1}_j(s)dM(s)\\
&=\phi({S}_j(\cdot); \tau)-\int^\tau_0[\int_s^\tau S_j(t)dt]Y^{-1}_j(s)dM(s)\\
\end{split}\]
where $I(\cdot)$ is the indicator function.\\

For a martingale $M(t)=N(t)-\Lambda(t)=N(t)-\int^t_0Y(s)dA(s)$, $K(t)$ a predictable function given history, $var(\int_0^tK(s)dM(s))=E(\int_0^tK^2(s)d\Lambda(s))$. Here we have $K_j(s)=(\int_s^\tau S_j(t)dt)Y^{-1}(s)$, and
\[
\begin{split}
var(\phi(\widehat{S}_j(\cdot)); \tau)
&=\int_0^\tau g_j^2(s, \tau)Y^{-2}_j(s)Y_j(s)dA_j(s)\\
&=\int_0^\tau g_j^2(s, \tau)Y^{-1}_j(s)dA_j(s)\\
&=n^{-1}\int_0^\tau g_j^2(s, \tau)\pi^{-1}_j(s)dA_j(s)\\
&=\zeta_j(\tau)\\
\end{split}\]
with $g_j(s, \tau)=-\int_s^\tau S_j(t)dt$. By the Martingale Central Limit Theorem, the process converges weakly to a gaussian process/Brownian motion.\\

Also, we have 
\[
\begin{split}
\phi(\widehat{S}_j(\cdot); \tau)-\phi(S_j(\cdot); \tau)
&=\int_0^\tau g_j(s, \tau)Y_j^{-1}(s)dM_j(s)\\
&=n^{-1}\int_0^\tau g_j(s, \tau)\pi_j^{-1}(s)dM_j(s)\\
&=n^{-1}\int_0^\tau g_j(s, \tau)\pi_j^{-1}(s)\sum_{i=1}^ndM_{ij}(s)\\
&=n^{-1}\sum_{i=1}^n\int_0^\tau g_j(s, \tau)\pi_j^{-1}(s)dM_{ij}(s)\\
&=n^{-1}\sum_{i=1}^n\Psi_{ij}(\tau)\\
&=n^{-1}\Psi_j(\tau),\\
\end{split}\]
where $M_{ij}(s)$ is the contribution to the total martingale process from subject $i$,$\Psi_{ij}(\tau)=\int_0^\tau g_j(s, \tau)\pi_j^{-1}(s)dM_{ij}(s)$ is the individual influence curve, and $\Psi_j(\tau)=\int_0^\tau g_j(s, \tau)\pi_j^{-1}(s)dM_j(s)$ the sum of influence curves. Note that $M_{ij}$ could be estimated at each discrete event times from the Kaplan-Meier curve. The covariance structure could then be estimated from the influence functions.\\

\begin{center}
{\large\bf SUPPLEMENTARY MATERIAL}
\end{center}

\begin{description}

\item[Web Appendix:]

\item[R code:] R code that implements the proposed method and performs simulation studies. (GNU zipped tar file)

\item[ACTT-1 data set:] The NIAID ACTT-1 trial data that supports the findings in this paper are publicly available from the website AccessClinicalData@NIAID. Researchers may apply for access to the data at \url{{https://accessclinicaldata.niaid.nih.gov/study-viewer/clinical_trials/ACTT}}.

\end{description}

\bigskip

\bibliographystyle{agsm}

\bibliography{bibliography}

\newpage

\begin{table}[h]
\caption{Sumamry Statistics of the estimators of RMSTs under two sample sizes based on 1000 replicates}
\begin{center}
\begin{tabular}{ccccccccccc} \hline
Parameter & Bias & SE & SEE & CP & Events & Bias & SE & SEE & CP & Events\\
 & & & N=100 & & & & & N=400 & &\\
$\tau=1$ & & & & & & & & & &\\
$RMST_1$ & 0.001& 0.035& 0.036& 0.955& 49.03& 0.000& 0.017& 0.018& 0.964& 196.34\\
$RMST_2$ & 0.001& 0.031& 0.030& 0.939& 34.48& 0.000& 0.015& 0.015& 0.964& 138.01\\
$RMST_3$ & 0.001& 0.029& 0.029& 0.930& 27.16& 0.000& 0.015& 0.014& 0.949& 109.02\\
$RMST_4$ & 0.000& 0.020& 0.019& 0.923& 11.49& 0.000& 0.010& 0.010& 0.937& 45.68\\
$\tau=1.5$ & & & & & & & & & &\\
$RMST_1$ & 0.002& 0.055& 0.056& 0.956& 59.28& 0.001& 0.027& 0.028& 0.961& 236.90\\
$RMST_2$ & 0.001& 0.052& 0.051& 0.938& 46.51& 0.001& 0.026& 0.026& 0.953& 185.11\\
$RMST_3$ & 0.002& 0.051& 0.050& 0.939& 37.20& 0.000& 0.026& 0.025& 0.951& 148.48\\
$RMST_4$ & 0.000& 0.038& 0.036& 0.926& 16.98& 0.000& 0.019& 0.018& 0.944& 67.40\\
$\tau=2$ & & & & & & & & & &\\
$RMST_1$ & 0.002& 0.072& 0.074& 0.962& 64.98& 0.001& 0.036& 0.037& 0.960& 259.47\\
$RMST_2$ & 0.001& 0.072& 0.072& 0.941& 53.78& 0.001& 0.036& 0.036& 0.954& 214.58\\
$RMST_3$ & 0.002& 0.073& 0.072& 0.948& 44.22& 0.001& 0.037& 0.036& 0.945& 176.60\\
$RMST_4$ & 0.000& 0.058& 0.056& 0.926& 21.52& 0.001& 0.029& 0.028& 0.935& 85.80\\

\hline
\vspace{-8mm}
\label{t1}
\end{tabular}
\end{center}
\end{table}

\newpage

\begin{table}[h]
\begin{center}
\caption{RMSTs at Day 20 in ACTT-1, point estimate with 95\% CI}
\begin{tabular}{cccc} 
 \hline
$\tau=20$ & Placebo & Remdesivir & Type II difference \\ \hline
 $RMST_1$ &  11.28(10.61, 11.95) & 13.10(12.44, 13.75) & 1.82(0.86, 2.78)\\ 
 $RMST_2$ &  17.36(16.90, 17.82) & 18.27(17.88, 18.65) & 0.91(0.24, 1.57)\\
 $RMST_3$ &  18.84(18.54, 19.14) & 19.36(19.13, 19.58) & 0.52(0.04, 1.00)\\
 \hline
\label{t2}
\end{tabular}
\end{center}
\end{table}

\newpage

\begin{figure}[h]
\begin{center}
\includegraphics[width=6in]{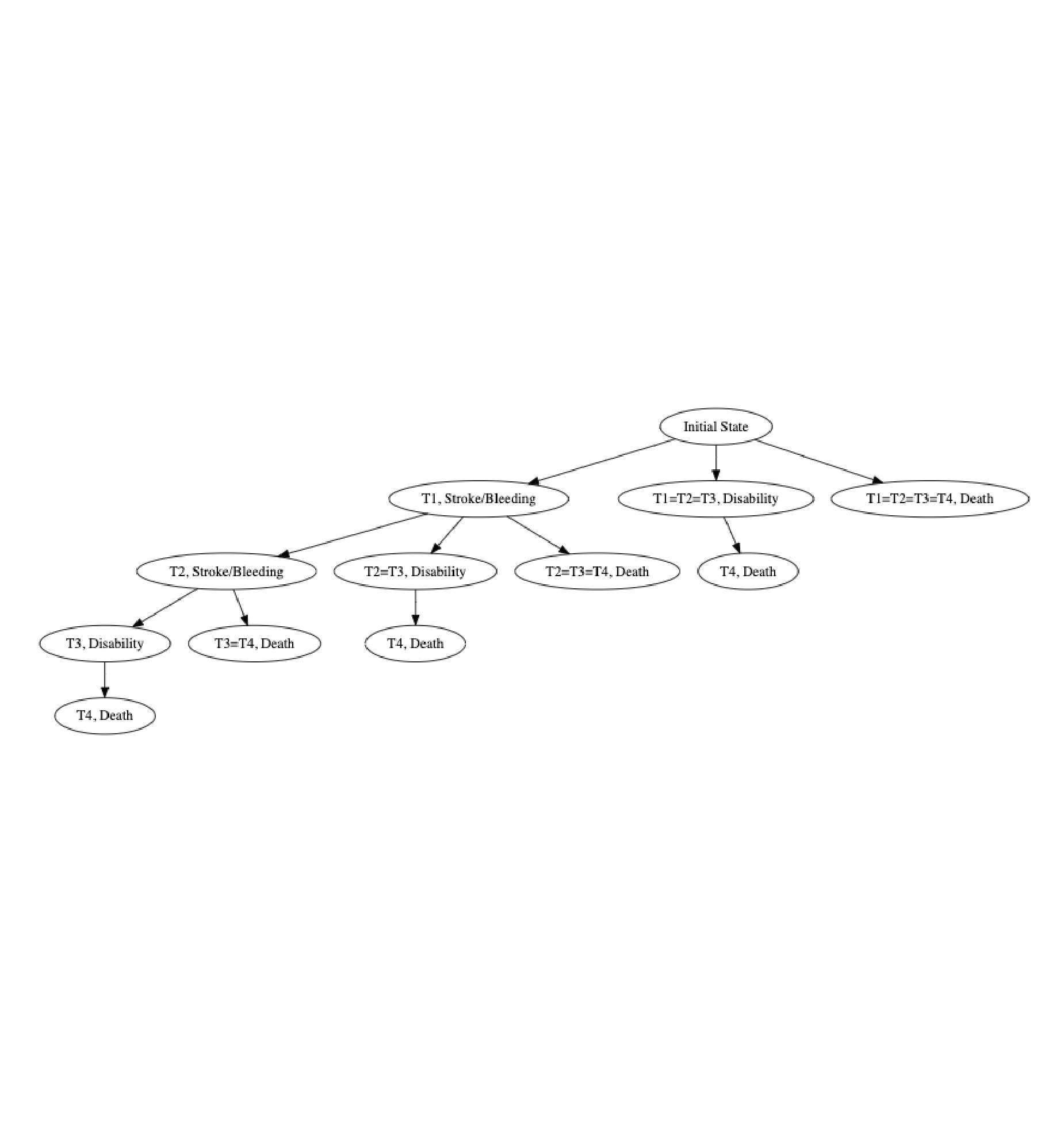}
\end{center}
\caption{Diagram of DOOR progression}
\label{fig1}
\end{figure}

\newpage

\begin{figure}[h]
\begin{subfigure}{.5\textwidth}
    \centering
    \includegraphics[width=\linewidth, height=6cm]{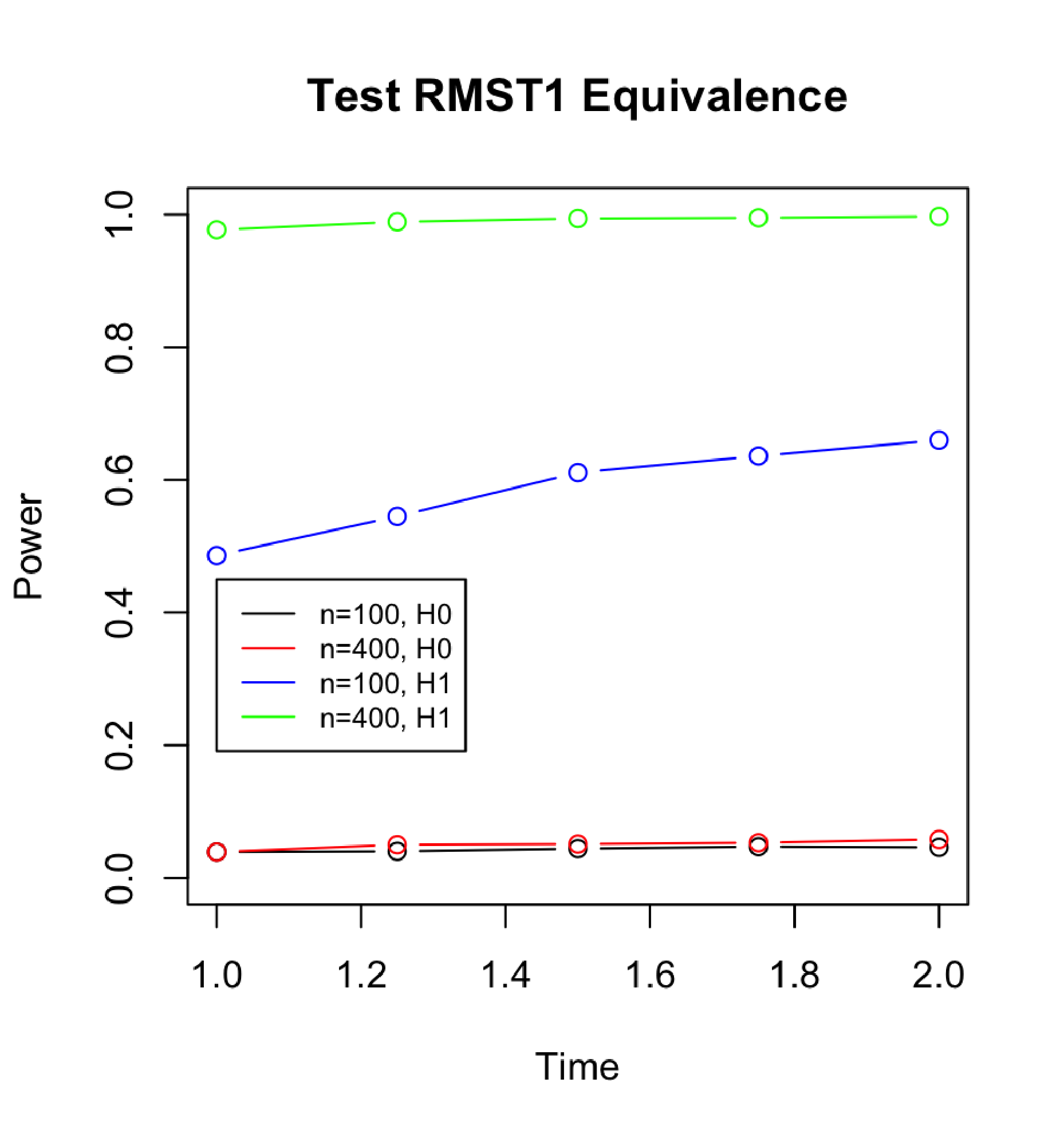}
\end{subfigure}%
\begin{subfigure}{.5\textwidth}
    \centering
    \includegraphics[width=\linewidth, height=6cm]{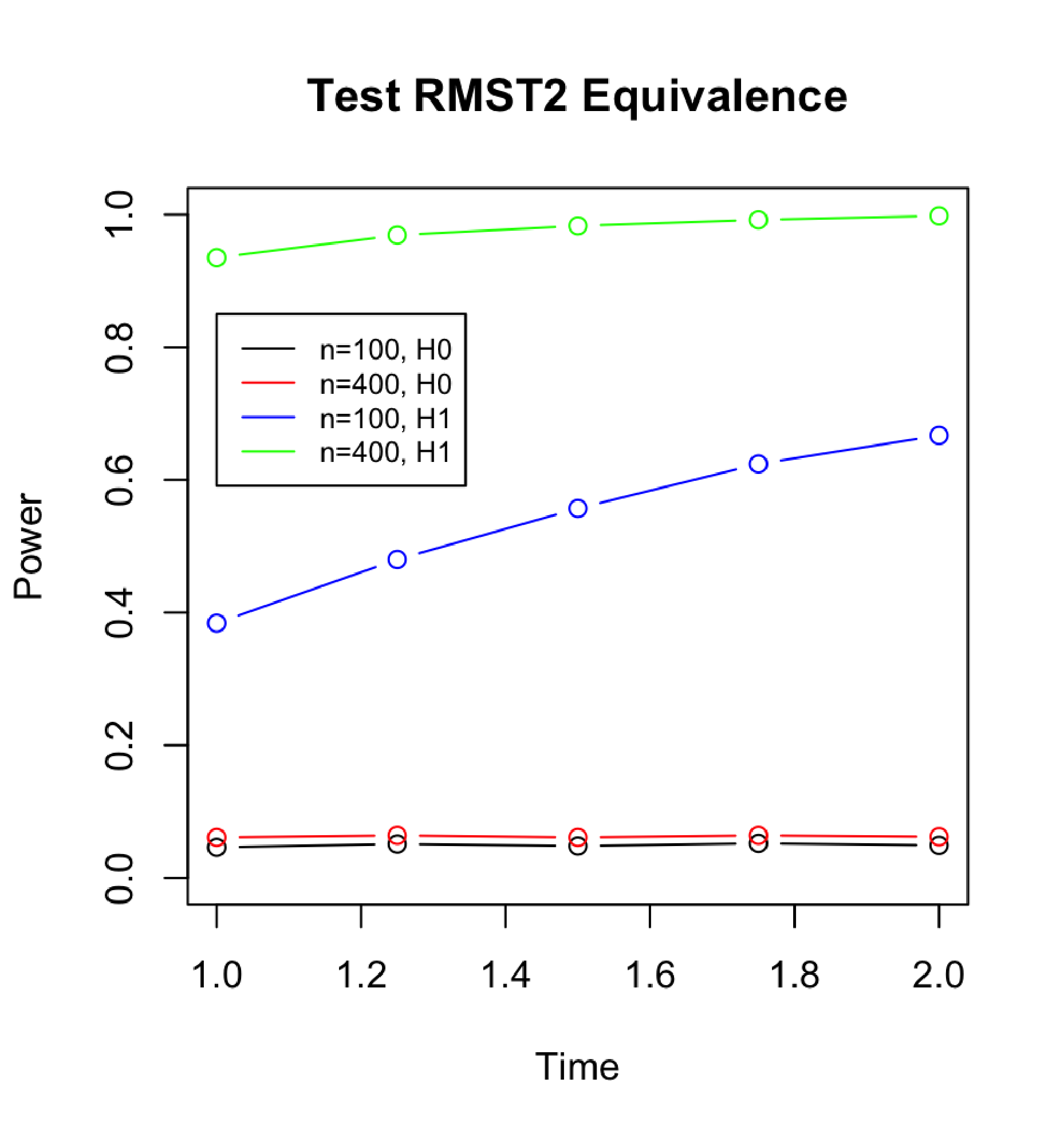}
\end{subfigure}
\begin{subfigure}{.5\textwidth}
    \centering
    \includegraphics[width=\linewidth, height=6cm]{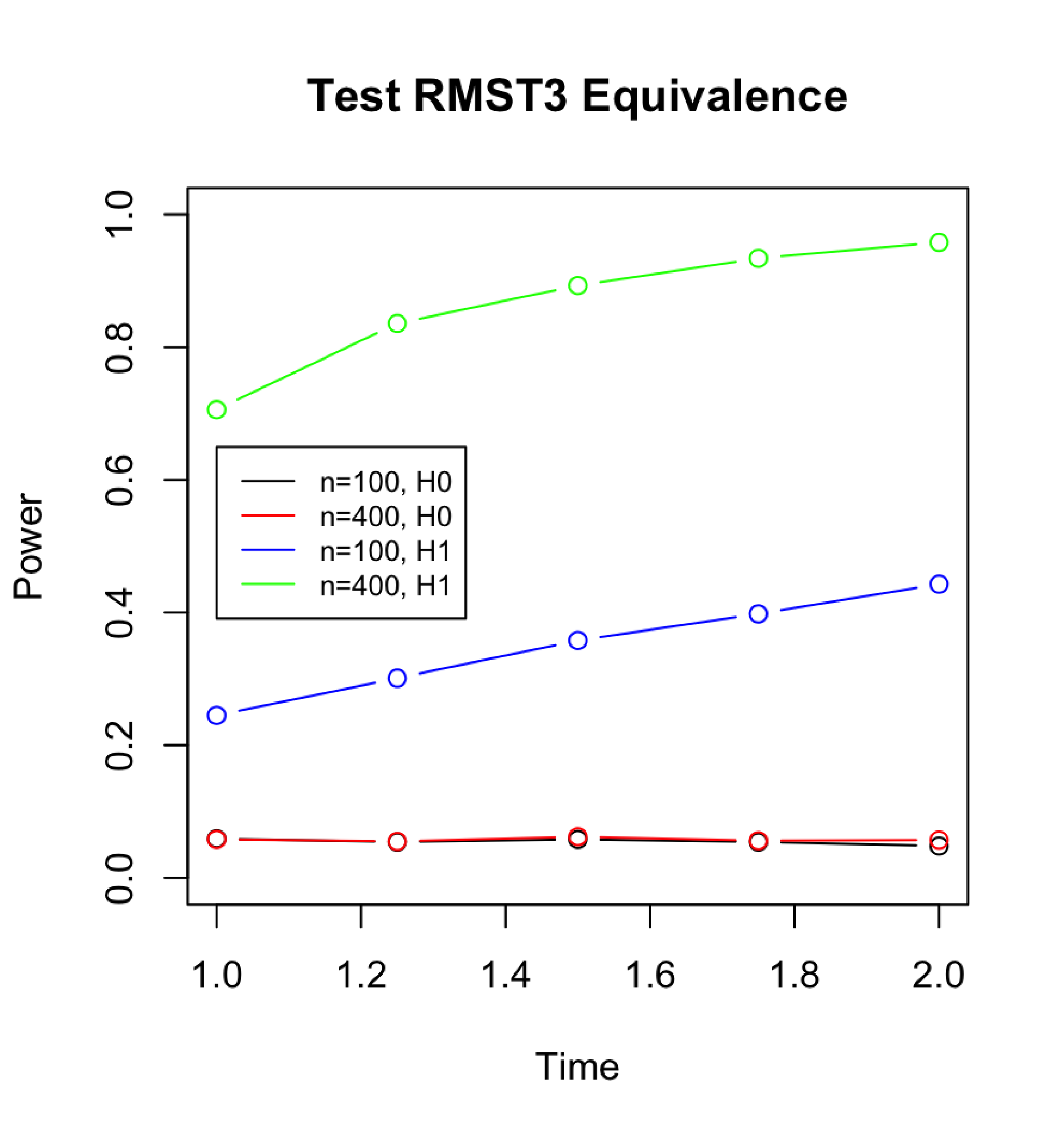}
\end{subfigure}%
\begin{subfigure}{.5\textwidth}
    \centering
    \includegraphics[width=\linewidth, height=6cm]{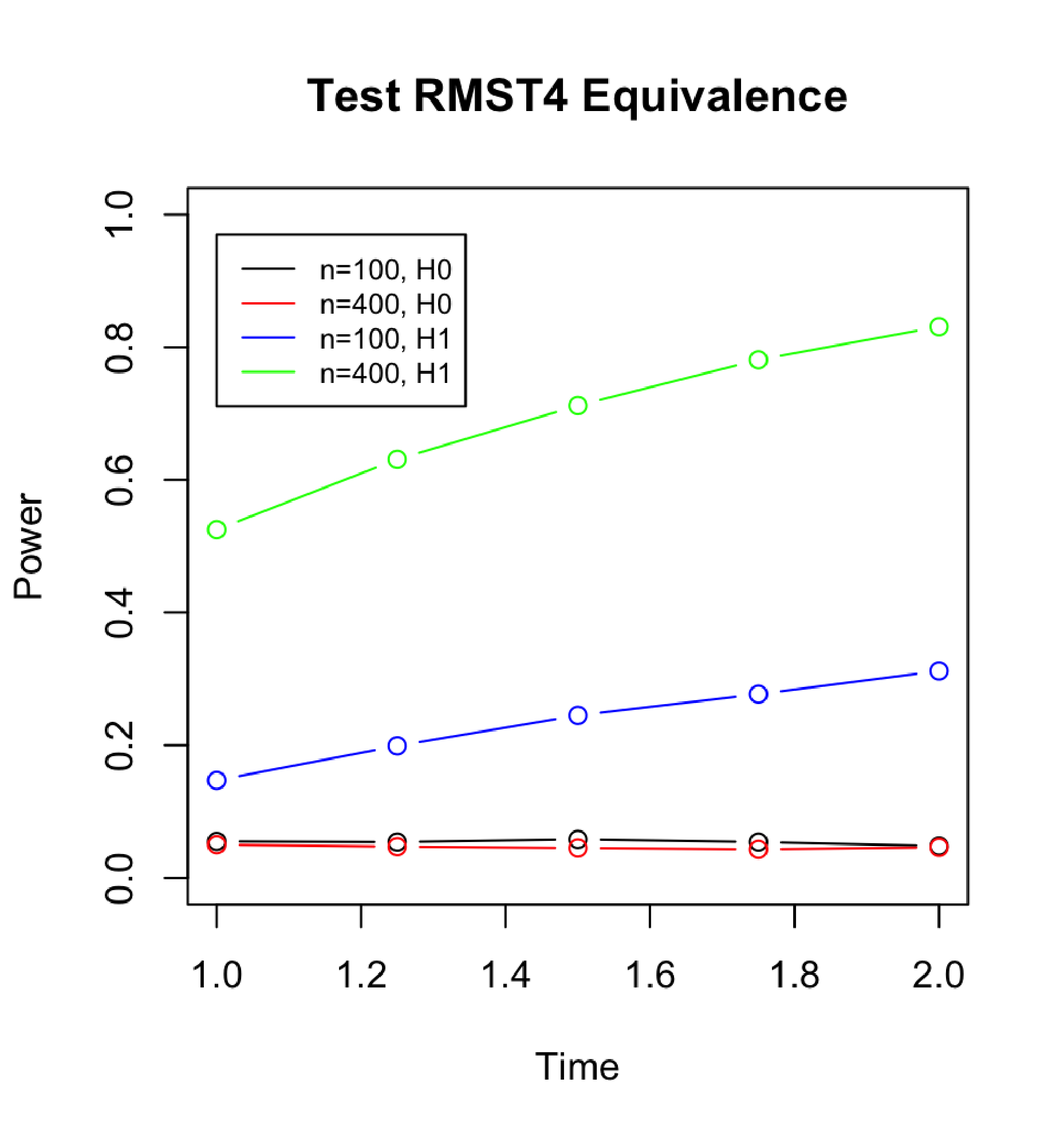}
\end{subfigure}
\caption{Type I error rates and Powers for Type II Inference: Testing the equivalence of individual level RMSTs between treatment and placebo}
\label{fig2}
\end{figure}

\newpage

\begin{figure}[h]
\begin{center}
\includegraphics[width=\textwidth, height=12cm]{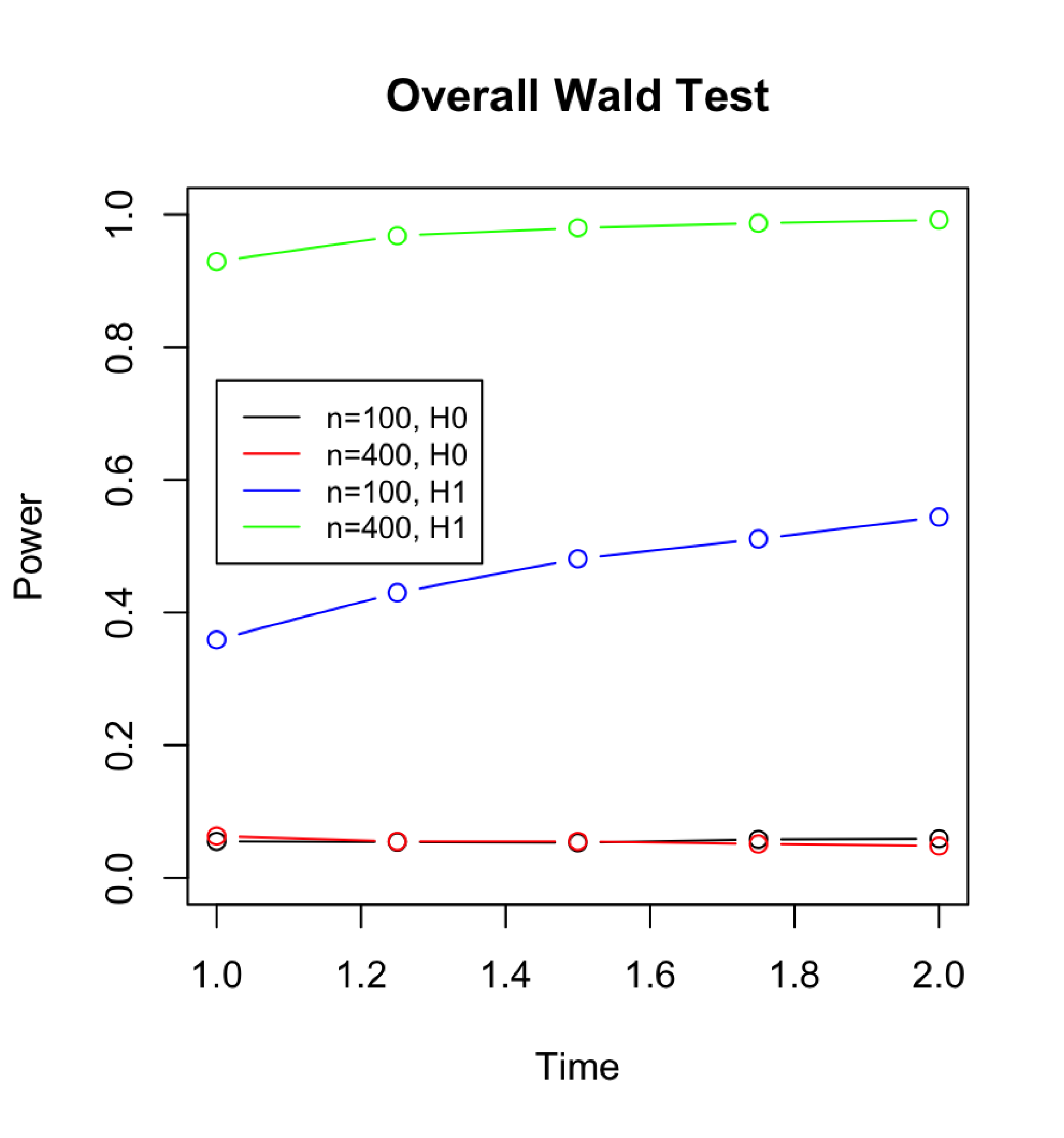}
\end{center}
\caption{Type I error rates and Powers for Type IV Inference: Testing the overall equivalence}
\label{fig3}
\end{figure}

\newpage

\begin{figure}[h]
\begin{center}
\includegraphics[width=\textwidth]{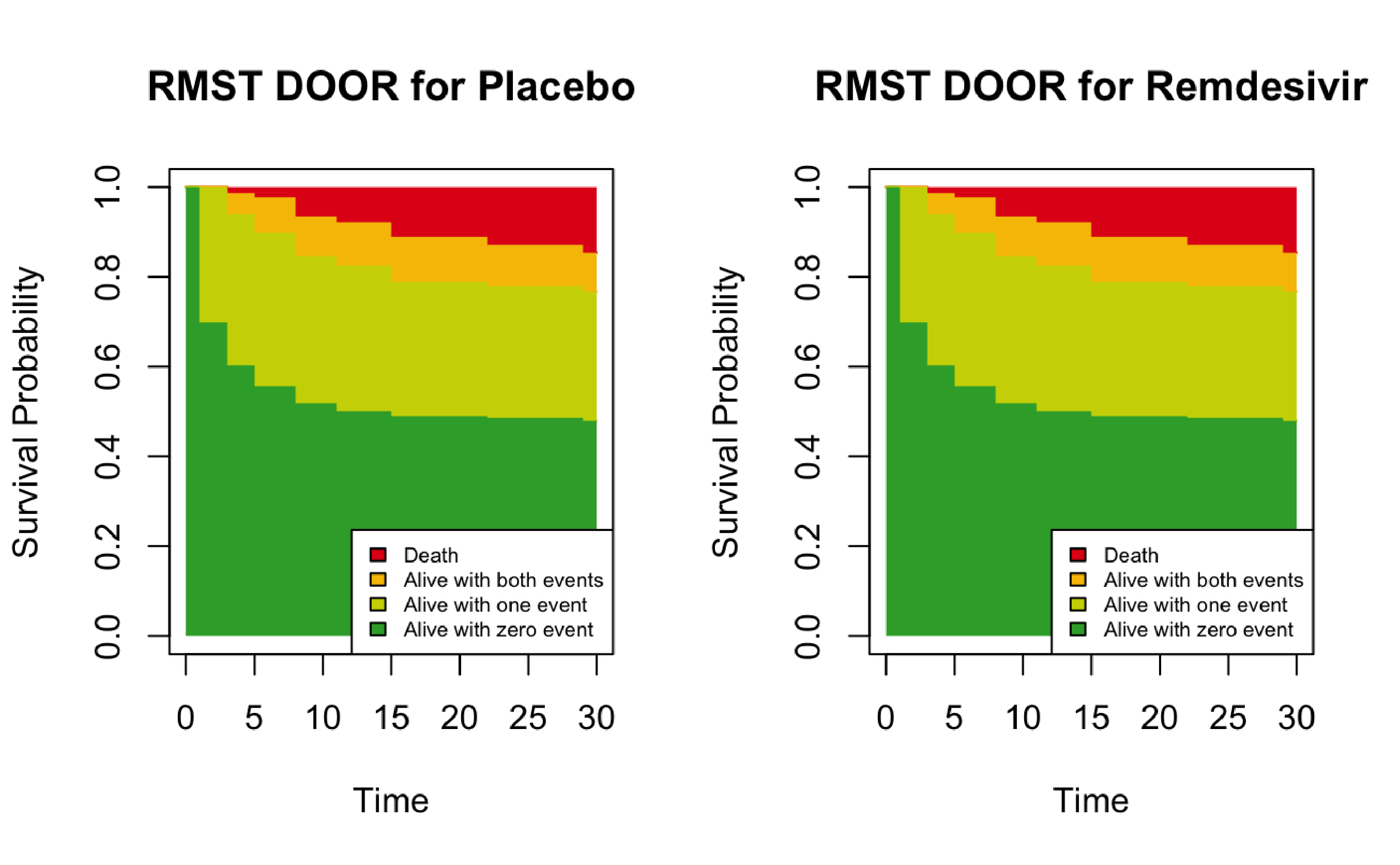}
\end{center}
\caption{RMST DOORs comparing Remdesivir and Placebo}
\label{fig4}
\end{figure}

\newpage

\begin{figure}[h]
\begin{center}
\includegraphics[width=\textwidth]{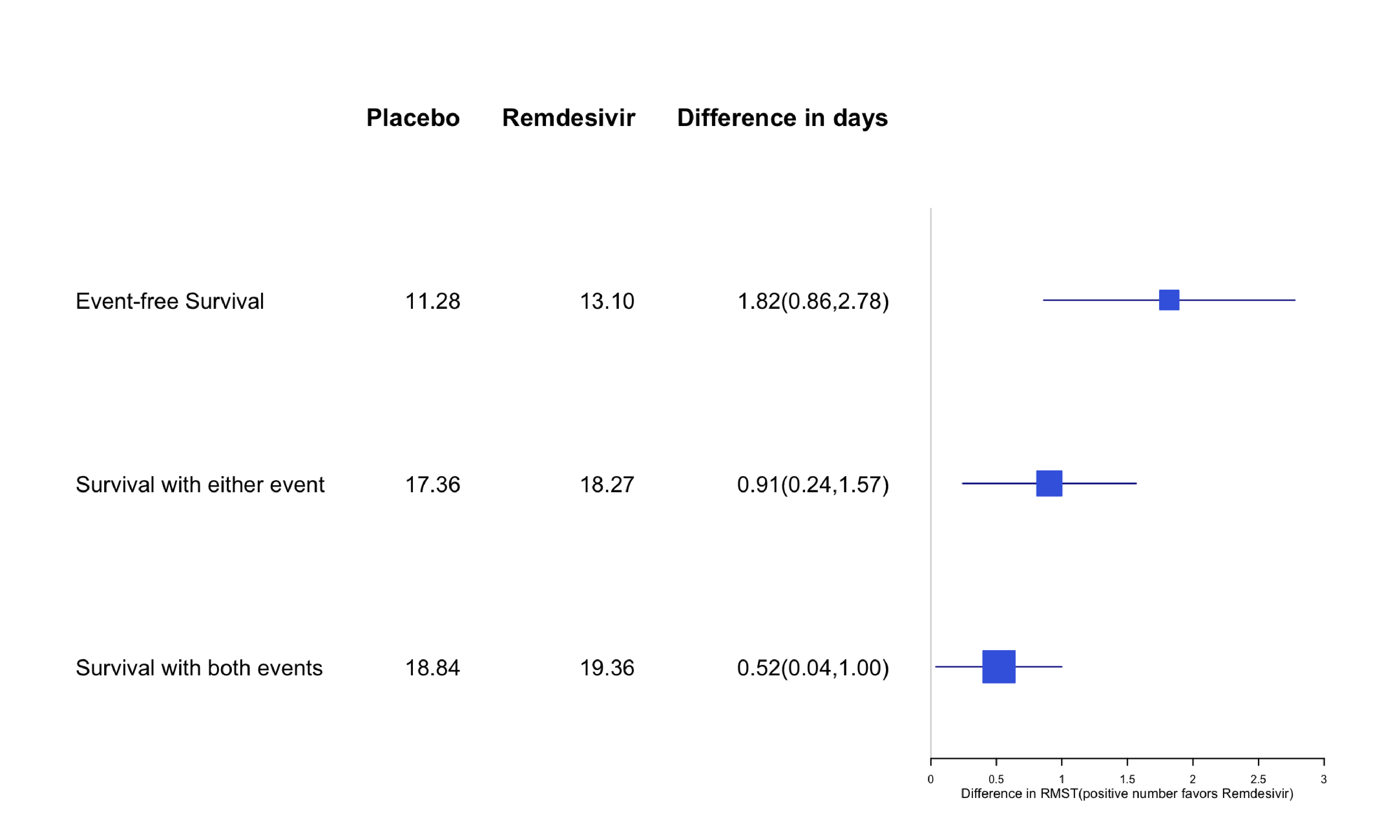}
\end{center}
\caption{RMSTs for tiered DOOR outcome for Remdesivir and Placebo}
\label{fig5}
\end{figure}

\end{document}